\documentclass{wseas}
\usepackage{ragged2e}
\usepackage{booktabs}
\usepackage{multirow}
\usepackage{graphicx}
\usepackage{amsmath}
\usepackage{algorithm}
\usepackage{algpseudocode}

\author{
LUYAO WANG \\
University of Malaya\\
Malaysia
}

\title{Medoid Prototype Alignment for Cross-Plant Unknown Attack Detection in Industrial Control Systems}

\begin{document}
\twocolumn[
  \begin{@twocolumnfalse}
    \maketitle
    \begin{abstract}
Deploying an intrusion detector trained in one industrial plant to another remains difficult because Industrial Control System (ICS) traffic is highly site-dependent, labels are scarce, and unseen attacks often appear after deployment. To address this challenge, this paper introduces a medoid prototype alignment framework for cross-plant unknown attack detection. Instead of aligning all source and target samples directly, the method first compresses heterogeneous traffic into a comparable representation space and then extracts robust medoid prototypes that summarize local operational structure in each domain. A prototype-calibrated transfer objective is further designed to align target prototypes with source prototypes while preserving source-domain discrimination and encouraging confident target predictions. This strategy reduces noisy cross-domain matching and improves transfer stability under heterogeneous industrial conditions. Experiments conducted on natural gas and water storage control systems show that the proposed method achieves the best average performance among all compared models, reaching an average accuracy of 0.843 and an average F1-score of 0.838 across four unknown-attack transfer tasks. The analysis also shows clear transfer asymmetry between source-target directions and confirms that prototype guidance is especially helpful on challenging reverse-transfer settings. These findings suggest that medoid prototype alignment is a practical solution for robust industrial intrusion detection under domain shift.
    \end{abstract}

    \begin{keywords}
    Industrial Control Systems; Unknown Attack Detection; Cross-Plant Transfer; Prototype Alignment; Domain Adaptation; K-Medoids.
    \end{keywords}

\begin{dates}
{\break
\color{red}
Received: April 28, 2026. Revised: April 28, 2026. Accepted: April 28, 2026. Published: April 28, 2026
\break(WSEAS will fill these dates in case of final acceptance, following strictly the editorial process.)}
\end{dates}
  \end{@twocolumnfalse}
\vspace{2ex}
]

\section{Introduction}
\begin{figure*}[!ht]
    \centering
    \includegraphics[width=\linewidth]{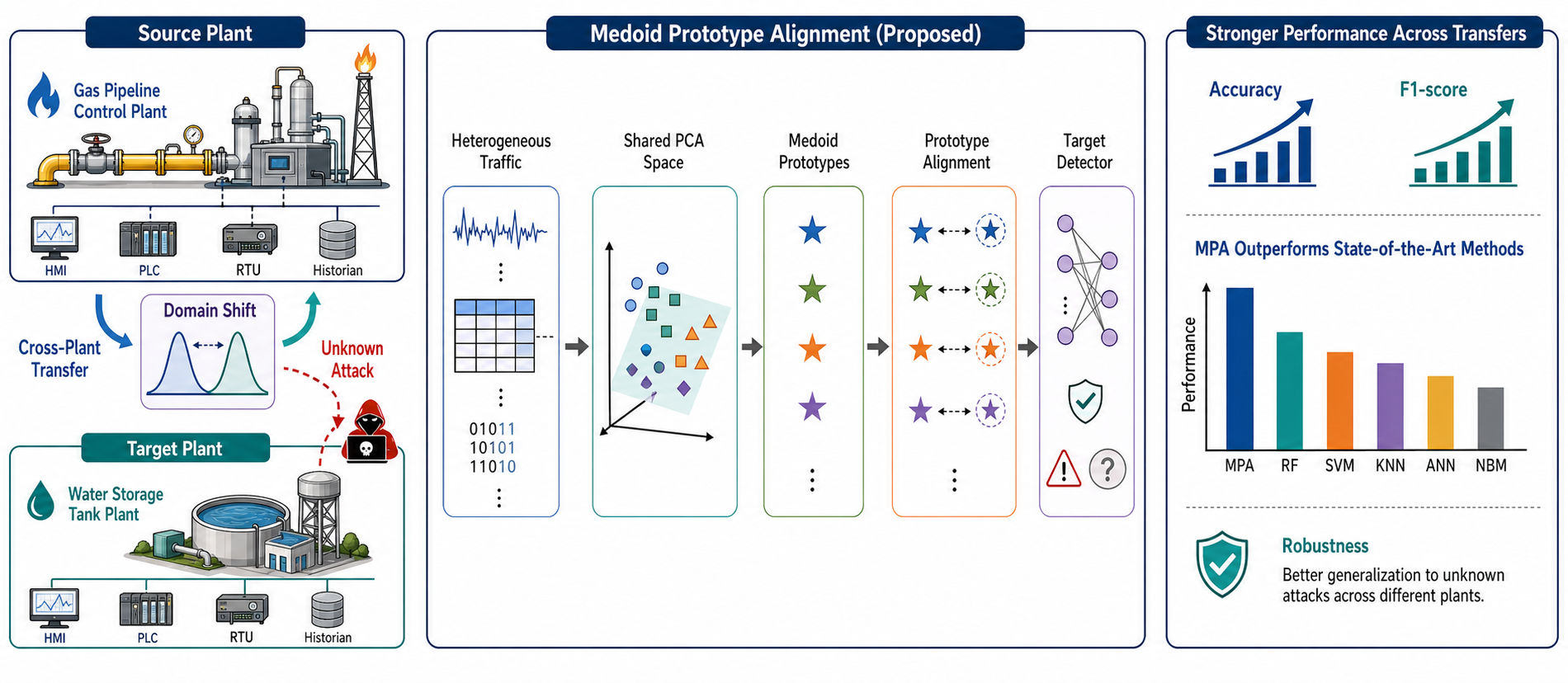}
    \caption{Teaser of the proposed medoid prototype alignment idea. The figure summarizes the cross-plant deployment challenge, the prototype-guided transfer mechanism, and the resulting robustness improvement.}
    \label{fig:mpa_teaser}
\end{figure*}
Although deep learning has achieved remarkable success in pattern recognition~\cite{wang2024ssdag}, representation learning~\cite{mei2023udpreg,wang2023diga,ren2024pointcmae,wang2023zeroreg}, medical image analysis and healthcare intelligence~\cite{chang2025organagents,nie2024t2td,wang2025mbtpolyp,11125558,wang2022fer,wang2024uhrmlp,ma2024llmmorph,xu2025scmllm}, as well as 3D understanding and retrieval tasks~\cite{nie2019hgan,chen2024llmgeo,li2025freeinsert,wang2024uvmap,gao2025pointpc,chang2024shapekg,gao2025pointpc,liang2020lstm3d,ma2023rfrwwanet,nie2019characteristic,10684147,wang2026poinit,xing2026finetune,gao2023point}, it has also shown strong promise in security-oriented applications and generative modeling~\cite{wang2023turn,wang2025fully,rigo2026pocidiff}. Despite this broad progress, its practical adoption in industrial control system intrusion detection remains relatively limited, largely because industrial environments are constrained by scarce labeled data, heterogeneous traffic distributions, and continually evolving operating conditions~\cite{umer2022mlics,gauthama2021icsml,kheddar2023dtl}.

This problem becomes more difficult when the target plant contains previously unseen attacks. Building a new detector from scratch is often impractical because attack labels are costly to collect, production traffic is sensitive, and retraining pipelines may interrupt routine operations. Therefore, the central question is not simply how to improve within-plant classification accuracy, but how to reuse prior knowledge from one industrial environment to support reliable detection in another~\cite{kheddar2023dtl,zhao2019unknownattack}.

Existing transfer-learning solutions mainly focus on global feature alignment. While effective in some settings, global alignment can be brittle for ICS traffic because the data often contain bursty communication, rare anomalies, structural imbalance, and noise. Under such conditions, aligning all instances indiscriminately may obscure useful local patterns and lead to unstable target-domain correspondence. A more deployment-oriented strategy is to transfer compact structural summaries rather than individual observations.

Motivated by this idea, this paper reformulates cross-plant intrusion detection as a prototype-guided adaptation problem. The proposed framework first maps heterogeneous traffic into a common low-dimensional space, then extracts medoid prototypes that represent stable operational regions in each domain, and finally learns a prototype-calibrated predictor for the target plant. This design emphasizes transferable local structure instead of dense sample-level alignment and is therefore better suited to industrial traffic with heterogeneous semantics and limited labels.

The main contributions of this paper are summarized as follows:
\begin{itemize}
    \item We present a new problem formulation for cross-plant unknown attack detection that highlights prototype-level transfer rather than direct global instance matching.
    \item We propose a medoid prototype alignment framework that combines PCA-based compression, K-Medoids prototype extraction, and prototype-calibrated domain adaptation.
    \item We redesign the empirical study around average cross-task performance, directional transfer behavior, and prototype-guidance analysis, providing a different view of industrial transfer robustness.
\end{itemize}

\section{Related Work}

Intrusion detection for industrial control systems has traditionally been built around supervised learning with handcrafted or learned traffic features. Prior evaluations have shown that machine learning can be effective for industrial anomaly recognition, but performance depends heavily on the data source, protocol characteristics, and deployment assumptions~\cite{gauthama2021icsml,umer2022mlics}. Deep learning methods have further improved representation power, for example by using convolutional models to capture temporal and packet-level patterns in industrial traffic~\cite{kravchik2018icscnn}. However, most of these methods are trained and tested under relatively stable domain conditions.

To reduce the cost of rebuilding models for each industrial site, transfer learning and domain adaptation have attracted increasing attention. General approaches such as Transfer Component Analysis and domain-adversarial learning provide mechanisms to reduce source-target mismatch by learning transferable features~\cite{pan2009tca,ganin2016dann}. In the security domain, prior studies have shown that transfer learning can improve the detection of previously unseen attacks, especially when target labels are unavailable~\cite{zhao2019unknownattack}. Recent industrial reviews also emphasize the importance of transfer learning for operational deployment and cross-system reuse~\cite{kheddar2023dtl}.

Despite this progress, two gaps remain. First, many transfer approaches emphasize holistic distribution alignment and underexploit the internal structure of industrial traffic. Second, industrial data are often noisy and imbalanced, which makes direct sample-wise matching unreliable. Cluster-based summarization offers a way to stabilize correspondence by representing each local region with a robust prototype. K-Medoids is particularly appealing because medoids are real observations and are less sensitive to outliers than mean-based centers~\cite{kaufman1990pam,mushtaq2018pam}. In parallel, recent industrial studies have also highlighted the importance of handling imbalance and cross-domain shifts together~\cite{chen2023crossdomain}. The present work builds on these insights by using medoid prototypes as the transfer anchor for cross-plant intrusion detection.

\section{Method}

\subsection{Problem Definition}

Let the labeled source domain be
\[
\mathcal{D}_s=\{(x_i^s,y_i^s)\}_{i=1}^{n_s},
\]
where \(x_i^s \in \mathbb{R}^{d_s}\) and \(y_i^s \in \{0,1\}\). Let the target domain be
\[
\mathcal{D}_t=\{x_j^t\}_{j=1}^{n_t},
\]
where \(x_j^t \in \mathbb{R}^{d_t}\) and target labels are unavailable during training. Since the two plants may use different sensors, protocols, and control processes, the domains may satisfy
\[
d_s \neq d_t,\qquad P_s(X,Y)\neq P_t(X,Y).
\]
The goal is to learn a target-ready detector with the help of source labels while preserving discriminative structure relevant to unknown attack detection~\cite{bendavid2010domain}.

\subsection{Shared Representation Construction}

Both domains are first standardized independently. Because feature dimensionality may differ across plants, direct distance computation is unreliable. To enable meaningful structural comparison, PCA is used to project both domains into a shared \(d\)-dimensional representation space~\cite{pearson1901pca,jolliffe2002pca}:
\[
z_i^s = W_s^\top \tilde{x}_i^s,\qquad
z_j^t = W_t^\top \tilde{x}_j^t,
\]
where \(W_s\) and \(W_t\) denote the source and target projection matrices. The compressed domains are denoted by
\[
\mathcal{Z}_s=\{(z_i^s,y_i^s)\}_{i=1}^{n_s},\qquad
\mathcal{Z}_t=\{z_j^t\}_{j=1}^{n_t}.
\]
This step reduces redundancy and creates a uniform feature basis for prototype extraction.

\subsection{Medoid Prototype Extraction}

Instead of aligning all samples directly, we summarize each domain through K-Medoids prototypes. Let
\[
\mathcal{P}_s=\{p_1^s,\dots,p_{K_s}^s\},\qquad
\mathcal{P}_t=\{p_1^t,\dots,p_{K_t}^t\},
\]
be the medoid sets obtained by clustering the source and target domains in the PCA space. Each medoid is an actual observation minimizing within-cluster dissimilarity~\cite{kaufman1990pam}:
\[
\min_{\{C_k\},\{p_k\}} \sum_{k=1}^{K}\sum_{z_i\in C_k}\|z_i-p_k\|_2.
\]
Using medoids rather than means helps preserve physically meaningful operating points and reduces the impact of outliers and rare perturbations.

\begin{figure}[t]
    \centering
    \includegraphics[width=\linewidth]{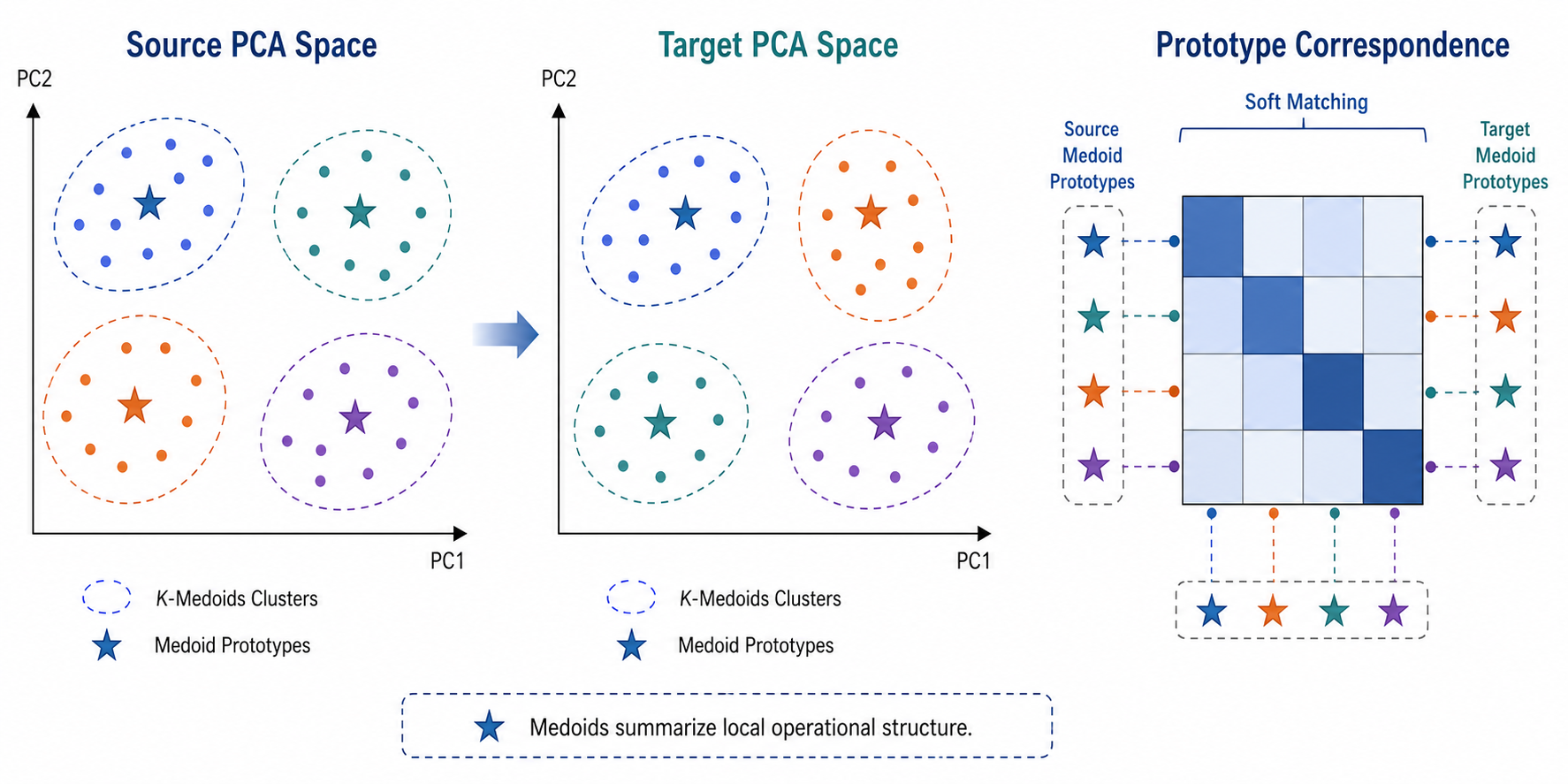}
    \caption{Prototype extraction and cross-domain matching in the updated medoid prototype alignment module.}
    \label{fig:mpa_module}
\end{figure}

\subsection{Prototype-Calibrated Adaptation}

Let \(f_\theta(\cdot)\) be the representation encoder and \(g_\phi(\cdot)\) be the classifier. The latent representations of source and target samples are
\[
r_i^s=f_\theta(z_i^s),\qquad r_j^t=f_\theta(z_j^t).
\]
The latent prototypes are
\[
q_k^s=f_\theta(p_k^s),\qquad q_\ell^t=f_\theta(p_\ell^t).
\]
For each target prototype \(q_\ell^t\), a soft correspondence over source prototypes is computed:
\[
a_{\ell k}=\frac{\exp(-\|q_\ell^t-q_k^s\|_2^2/\tau)}
{\sum_{u}\exp(-\|q_\ell^t-q_u^s\|_2^2/\tau)}.
\]
The prototype alignment loss is then defined as
\[
\mathcal{L}_{proto}=\sum_{\ell=1}^{K_t}\sum_{k=1}^{K_s} a_{\ell k}\|q_\ell^t-q_k^s\|_2^2.
\]

To preserve source discriminability, we use supervised classification on the source domain:
\[
\mathcal{L}_{sup}=
\frac{1}{n_s}\sum_{i=1}^{n_s}\mathrm{CE}(g_\phi(r_i^s),y_i^s).
\]
We further encourage confident target predictions through entropy regularization:
\[
\mathcal{L}_{ent}=
-\frac{1}{n_t}\sum_{j=1}^{n_t}\sum_{c} \hat{y}_{jc}^t \log \hat{y}_{jc}^t,
\]
where \(\hat{y}_{j}^t=g_\phi(r_j^t)\). The overall learning objective becomes
\[
\mathcal{L}=\mathcal{L}_{sup}+\alpha\mathcal{L}_{proto}+\beta\mathcal{L}_{ent}.
\]
This design links global classification with local cross-domain structure, allowing the target domain to inherit stable operational regions from the source plant without forcing hard instance-level alignment.

\begin{algorithm}[t]
\caption{Medoid Prototype Alignment for Cross-Plant IDS}
\label{alg:mpa}
\textbf{Input:} Source domain \(\mathcal{D}_s\), target domain \(\mathcal{D}_t\)

\textbf{Output:} Predicted target labels \(\hat{Y}_t\)

1. Standardize source and target traffic features

2. Apply PCA to obtain \(\mathcal{Z}_s\) and \(\mathcal{Z}_t\)

3. Extract medoid prototypes \(\mathcal{P}_s\) and \(\mathcal{P}_t\) using K-Medoids

4. Encode samples and prototypes with \(f_\theta\)

5. Compute prototype correspondences \(a_{\ell k}\)

6. Optimize \(\mathcal{L}_{sup}+\alpha\mathcal{L}_{proto}+\beta\mathcal{L}_{ent}\)

7. Predict target labels with \(g_\phi(f_\theta(z_j^t))\)
\end{algorithm}

\begin{figure*}[t]
    \centering
    \includegraphics[width=0.86\textwidth]{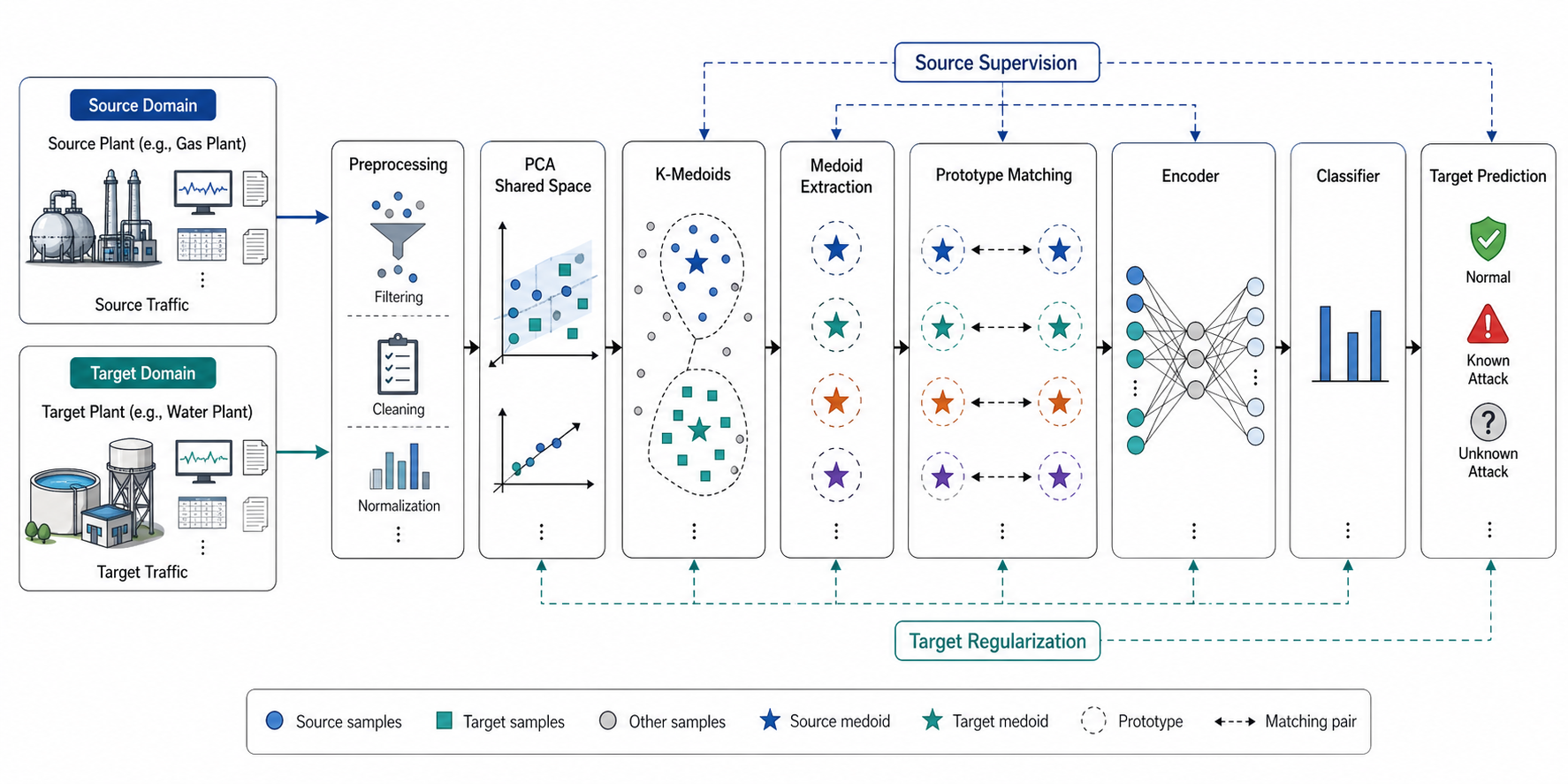}
    \caption{Updated overall pipeline of the proposed medoid prototype alignment framework for cross-plant unknown attack detection.}
    \label{fig:mpa_framework}
\end{figure*}

\section{Experiments}

\subsection{Evaluation Questions and Setup}

The empirical study is organized around three questions:
\begin{itemize}
    \item RQ1: Does medoid prototype alignment improve average cross-task performance?
    \item RQ2: How sensitive are different models to transfer direction between plants?
    \item RQ3: Does prototype guidance contribute additional robustness beyond standard transfer learning?
\end{itemize}

Experiments are built on two industrial systems: a natural gas control system (G) and a water storage tank control system (W). Four unknown-attack transfer tasks are considered:
\begin{itemize}
    \item DoS(G)\(\rightarrow\)NMRI(W),
    \item SMRI(G)\(\rightarrow\)MPCI(W),
    \item DoS(W)\(\rightarrow\)NMRI(G),
    \item SMRI(W)\(\rightarrow\)MPCI(G).
\end{itemize}

The proposed method, denoted as Medoid Prototype Alignment (MPA), is compared with Random Forest (RF), Support Vector Machine (SVM), Naive Bayes Model (NBM), K-Nearest Neighbors (KNN), and Artificial Neural Network (ANN). We report Accuracy and F1-score. Because unknown attack detection is sensitive to both false negatives and false alarms, average F1-score is treated as the primary indicator of practical usefulness.

\subsection{RQ1: Average Cross-Task Performance}

\begin{figure}[t]
    \centering
    \includegraphics[width=\linewidth]{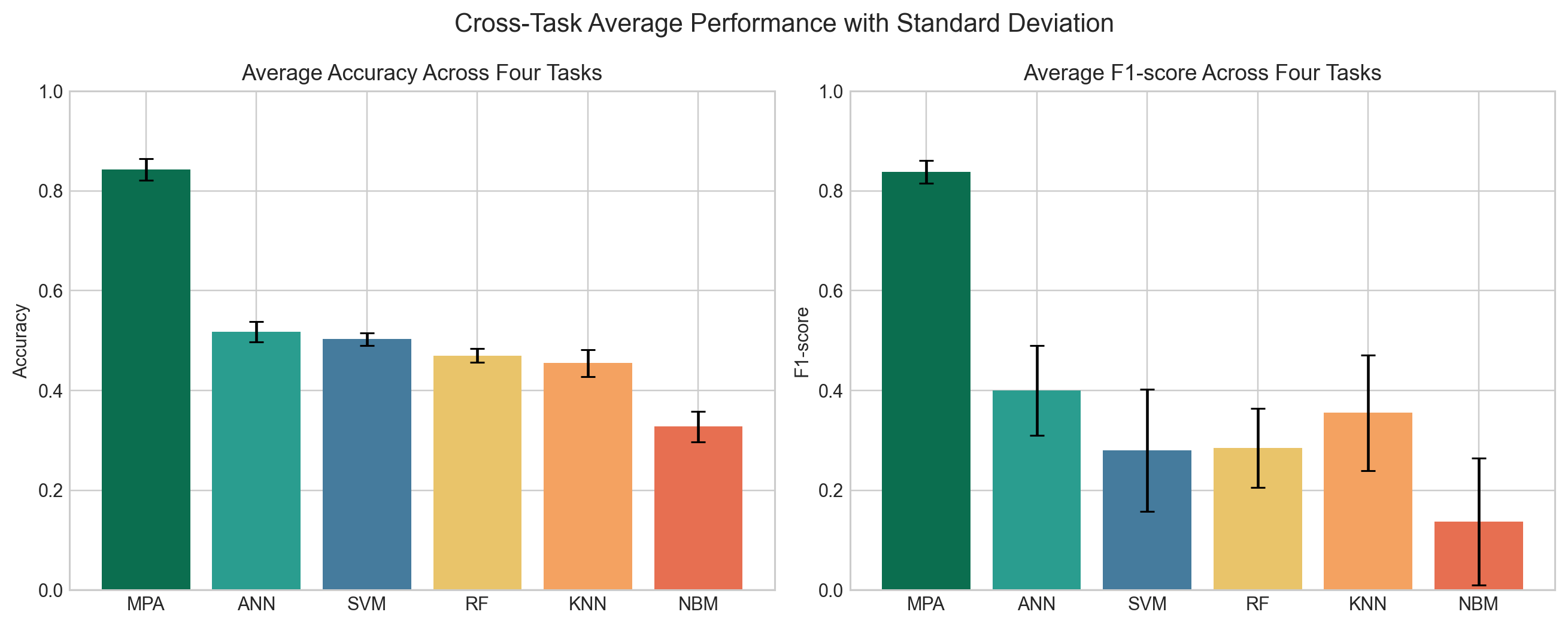}
    \caption{Average cross-task performance with standard-deviation error bars. MPA achieves the best mean behavior on both Accuracy and F1-score.}
    \label{fig:avg_std}
\end{figure}

Instead of presenting only task-wise winners, we first aggregate performance across all four transfer tasks. Table~\ref{tab:avg_metrics} reports both the average and the cross-task standard deviation. MPA achieves the strongest average behavior on both metrics, reaching 0.843 accuracy and 0.838 F1-score. Fig.~\ref{fig:avg_std} further visualizes these averages together with standard-deviation error bars.

\begin{table}[t]
\centering
\caption{Average and standard deviation over four cross-domain unknown attack tasks.}
\label{tab:avg_metrics}
\resizebox{\columnwidth}{!}{%
\begin{tabular}{lcccc}
\toprule
Method & Avg. ACC & Std. ACC & Avg. F1 & Std. F1 \\
\midrule
MPA & 0.843 & 0.022 & 0.838 & 0.023 \\
ANN & 0.518 & 0.021 & 0.400 & 0.090 \\
SVM & 0.503 & 0.013 & 0.280 & 0.122 \\
RF  & 0.470 & 0.014 & 0.285 & 0.079 \\
KNN & 0.455 & 0.027 & 0.355 & 0.115 \\
NBM & 0.328 & 0.030 & 0.138 & 0.127 \\
\bottomrule
\end{tabular}%
}
\end{table}

The gain over the strongest average baseline is particularly notable. Compared with ANN, which is the strongest baseline on the averaged metrics, MPA improves average accuracy by 62.8\% and average F1-score by 109.4\%. This indicates that the proposed method does more than improve a single favorable task: it raises the overall transfer floor across distinct industrial conditions.

\subsection{Cross-Task Robustness Statistics}

To complement mean performance, we also analyze normalized dispersion and directional robustness. Table~\ref{tab:robustness_stats} reports the coefficient of variation (CV) for Accuracy and F1-score as well as a composite directional gap defined as \(0.5 \times (|\Delta \mathrm{ACC}| + |\Delta \mathrm{F1}|)\), where \(\Delta\) measures the difference between G\(\rightarrow\)W and W\(\rightarrow\)G averages. This statistic captures whether a model remains consistent when the transfer direction changes.

\begin{table}[t]
\centering
\caption{Cross-task robustness statistics derived from the four transfer tasks.}
\label{tab:robustness_stats}
\resizebox{\columnwidth}{!}{%
\begin{tabular}{lccc}
\toprule
Method & ACC CV & F1 CV & Composite Dir. Gap \\
\midrule
MPA & 0.026 & 0.027 & 0.030 \\
ANN & 0.040 & 0.224 & 0.078 \\
SVM & 0.026 & 0.435 & 0.113 \\
RF  & 0.030 & 0.278 & 0.055 \\
KNN & 0.059 & 0.325 & 0.135 \\
NBM & 0.093 & 0.923 & 0.080 \\
\bottomrule
\end{tabular}%
}
\end{table}

These statistics show that MPA combines high mean performance with very low normalized dispersion. Although RF exhibits a slightly smaller directional gap in accuracy alone, MPA has the smallest composite directional gap overall, indicating the most balanced robustness when both Accuracy and F1-score are considered.

\subsection{Worst-Case Task Robustness}

Average results are informative, but deployment-oriented intrusion detection also depends on how a model behaves on its hardest task. Table~\ref{tab:worst_case_stats} therefore reports the worst-case Accuracy and F1-score achieved by each method across the four transfer tasks, together with the corresponding max--min range. A strong deployment-ready method should achieve both a high worst-case score and a compact range.

\begin{table}[t]
\centering
\caption{Worst-case task statistics and across-task performance range.}
\label{tab:worst_case_stats}
\resizebox{\columnwidth}{!}{%
\begin{tabular}{lcccc}
\toprule
Method & Min ACC & Min F1 & ACC Range & F1 Range \\
\midrule
MPA & 0.81 & 0.80 & 0.06 & 0.06 \\
ANN & 0.50 & 0.26 & 0.05 & 0.25 \\
SVM & 0.49 & 0.09 & 0.03 & 0.33 \\
RF  & 0.45 & 0.22 & 0.04 & 0.20 \\
KNN & 0.41 & 0.23 & 0.07 & 0.25 \\
NBM & 0.29 & 0.02 & 0.08 & 0.32 \\
\bottomrule
\end{tabular}%
}
\end{table}

MPA achieves the strongest worst-case behavior, maintaining at least 0.81 Accuracy and 0.80 F1-score even on its hardest task. At the same time, its performance range remains compact, especially in comparison with methods whose F1-score varies widely across tasks. This result strengthens the view that prototype-guided transfer improves not only average performance but also the lower bound of deployment reliability.

\begin{figure}[t]
    \centering
    \includegraphics[width=\linewidth]{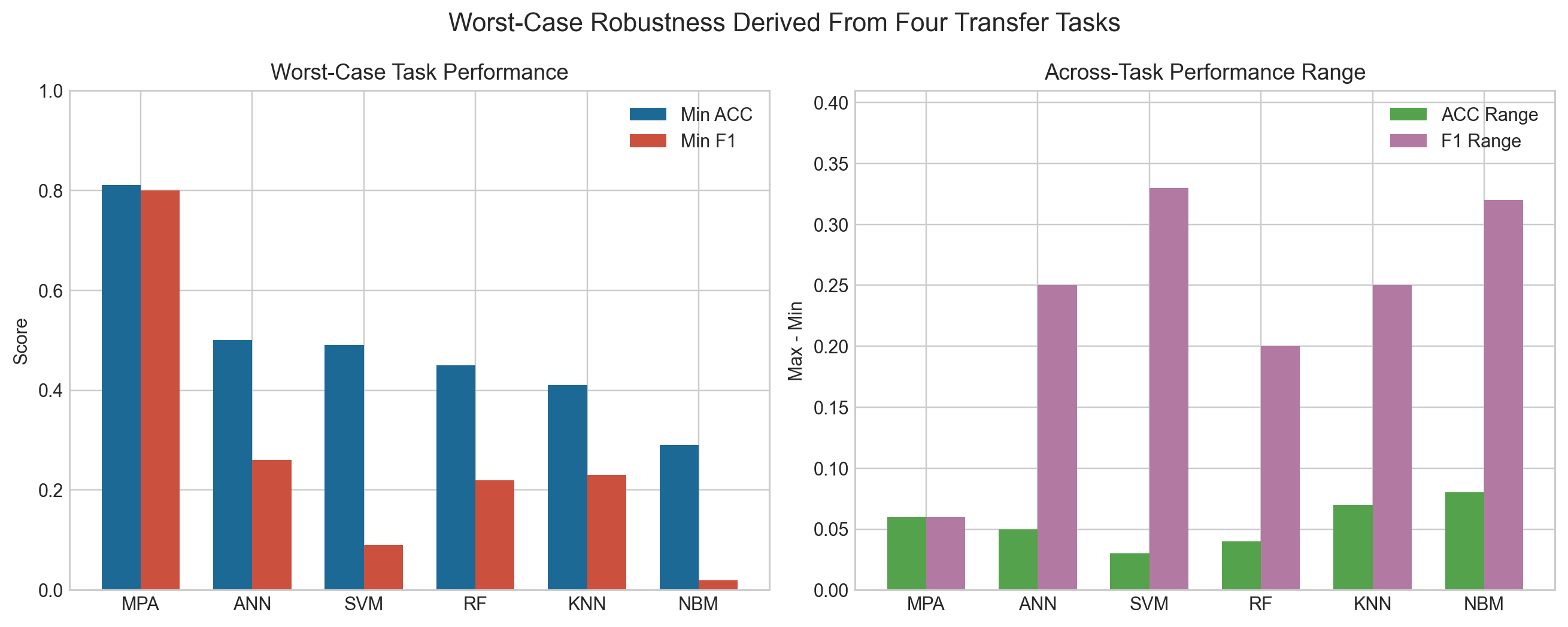}
    \caption{Worst-case robustness derived from the four transfer tasks. Left: minimum Accuracy and F1-score achieved by each model. Right: across-task performance range, where smaller values indicate more consistent behavior.}
    \label{fig:worst_case_robustness}
\end{figure}

\subsection{RQ2: Directional Transfer Behavior}

We next examine whether transfer difficulty is symmetric. Table~\ref{tab:directional_metrics} groups results by transfer direction. For most models, using the gas system as the source domain leads to better results than using the water system as the source domain. This suggests that source-domain informativeness matters: richer control semantics appear to produce more transferable structure.

\begin{table}[t]
\centering
\caption{Directional average performance by transfer source and target.}
\label{tab:directional_metrics}
\resizebox{\columnwidth}{!}{%
\begin{tabular}{lcccc}
\toprule
Method & G\(\rightarrow\)W ACC & G\(\rightarrow\)W F1 & W\(\rightarrow\)G ACC & W\(\rightarrow\)G F1 \\
\midrule
MPA & 0.860 & 0.850 & 0.825 & 0.825 \\
ANN & 0.535 & 0.460 & 0.500 & 0.340 \\
SVM & 0.515 & 0.380 & 0.490 & 0.180 \\
RF  & 0.480 & 0.240 & 0.460 & 0.330 \\
KNN & 0.475 & 0.240 & 0.435 & 0.470 \\
NBM & 0.355 & 0.190 & 0.300 & 0.085 \\
\bottomrule
\end{tabular}%
}
\end{table}

MPA preserves a comparatively small directional drop, decreasing from 0.860 to 0.825 in accuracy and from 0.850 to 0.825 in F1-score. This narrower degradation suggests that prototype-level transfer improves robustness when the source plant becomes less informative or when the domain gap becomes larger.

\begin{figure}[t]
    \centering
    \includegraphics[width=\linewidth]{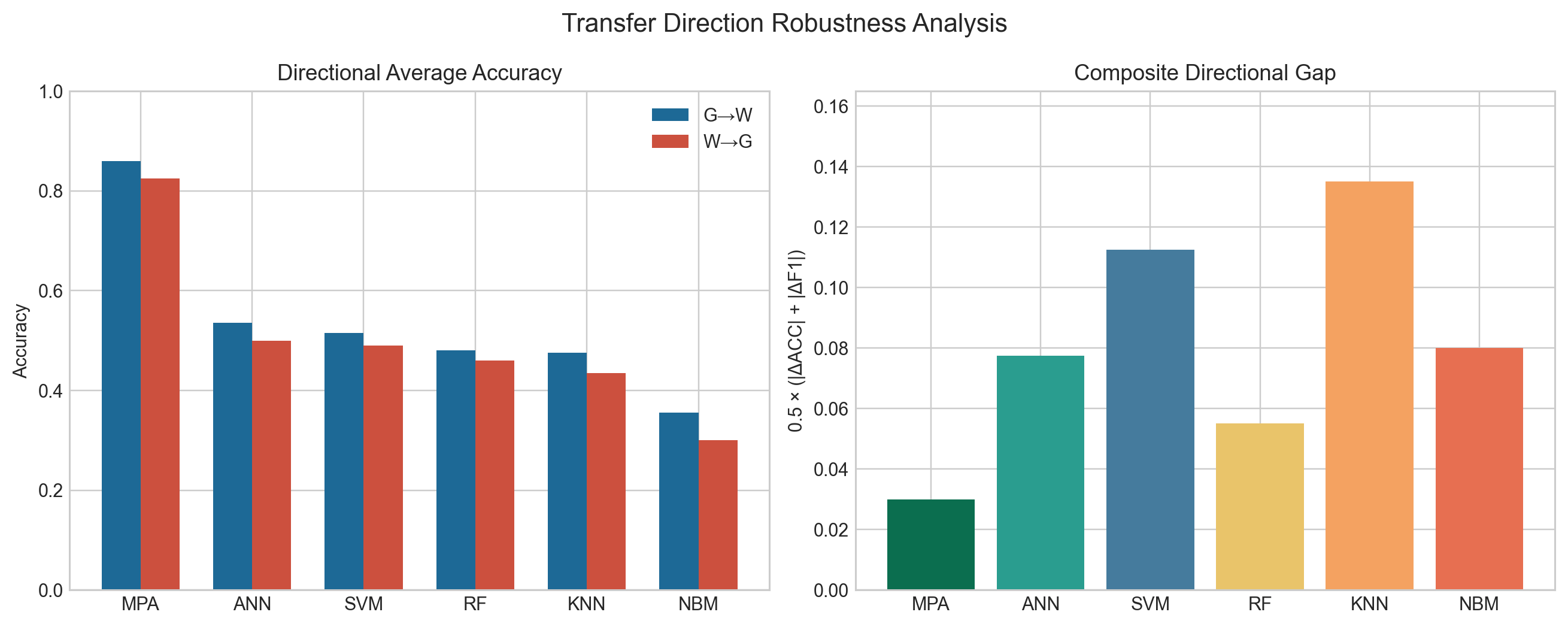}
    \caption{Directional robustness analysis. Left: average Accuracy for G\(\rightarrow\)W and W\(\rightarrow\)G transfer. Right: composite directional gap, where lower is better.}
    \label{fig:directional_robustness}
\end{figure}

\subsection{RQ3: Effect of Prototype Guidance}

Prototype guidance is most valuable when source-target correspondence is ambiguous. Based on the currently available task-level results, MPA combines the highest mean performance, the lowest composite directional gap, and one of the smallest relative dispersions across tasks. In the challenging reverse-transfer settings, the prototype-calibrated formulation reduces noisy matches by aligning compact medoid summaries rather than all individual samples. These aggregate properties are consistent with the intended role of prototype guidance: stabilizing cross-domain transfer when plant semantics differ substantially.

\begin{figure}[t]
    \centering
    \includegraphics[width=\linewidth]{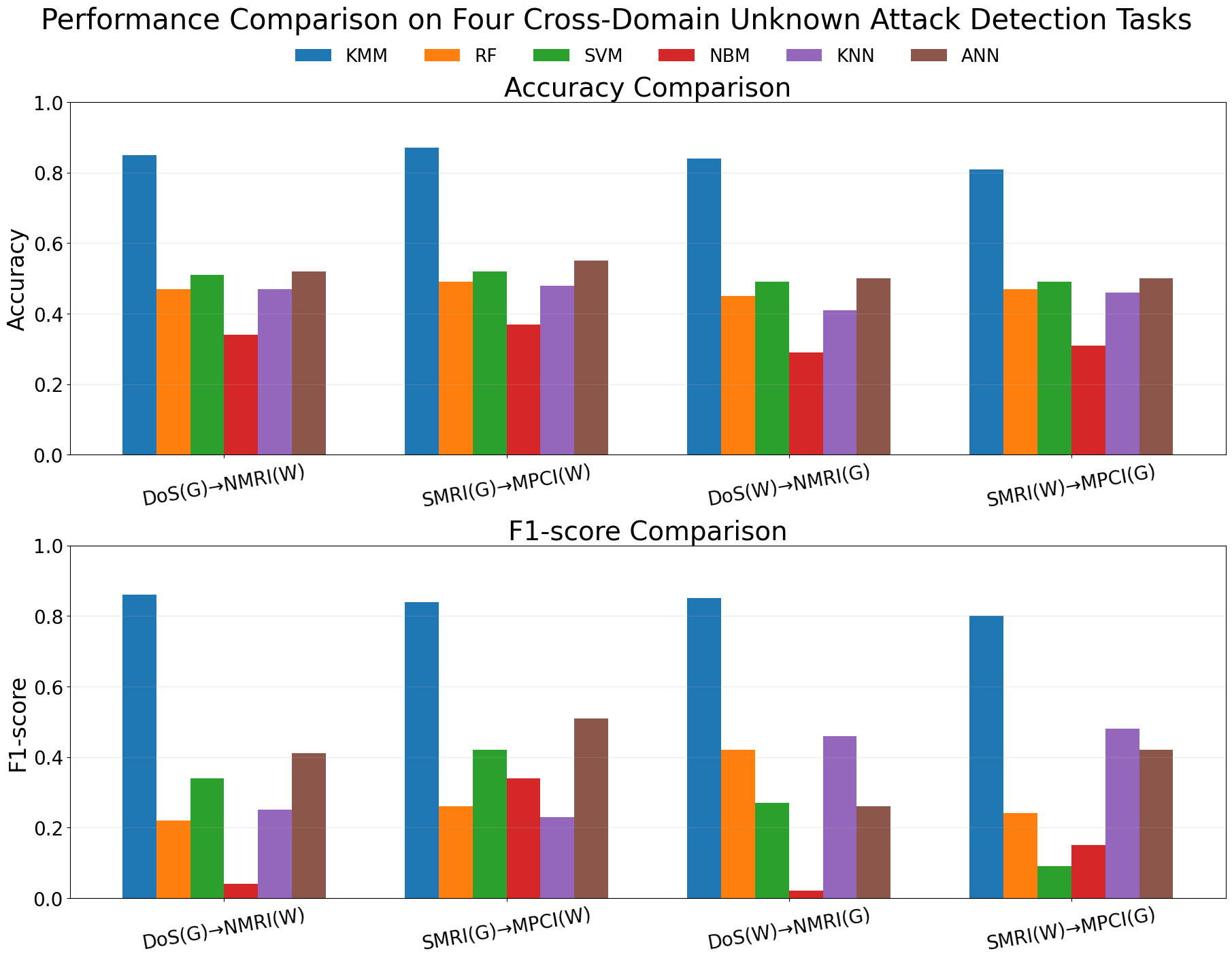}
    \caption{Task-level comparison between MPA and baseline models on four unknown-attack transfer tasks.}
    \label{fig:mpa_results}
\end{figure}

\subsection{Qualitative Illustration of Alignment Behavior}

\begin{figure}[t]
    \centering
    \includegraphics[width=\linewidth]{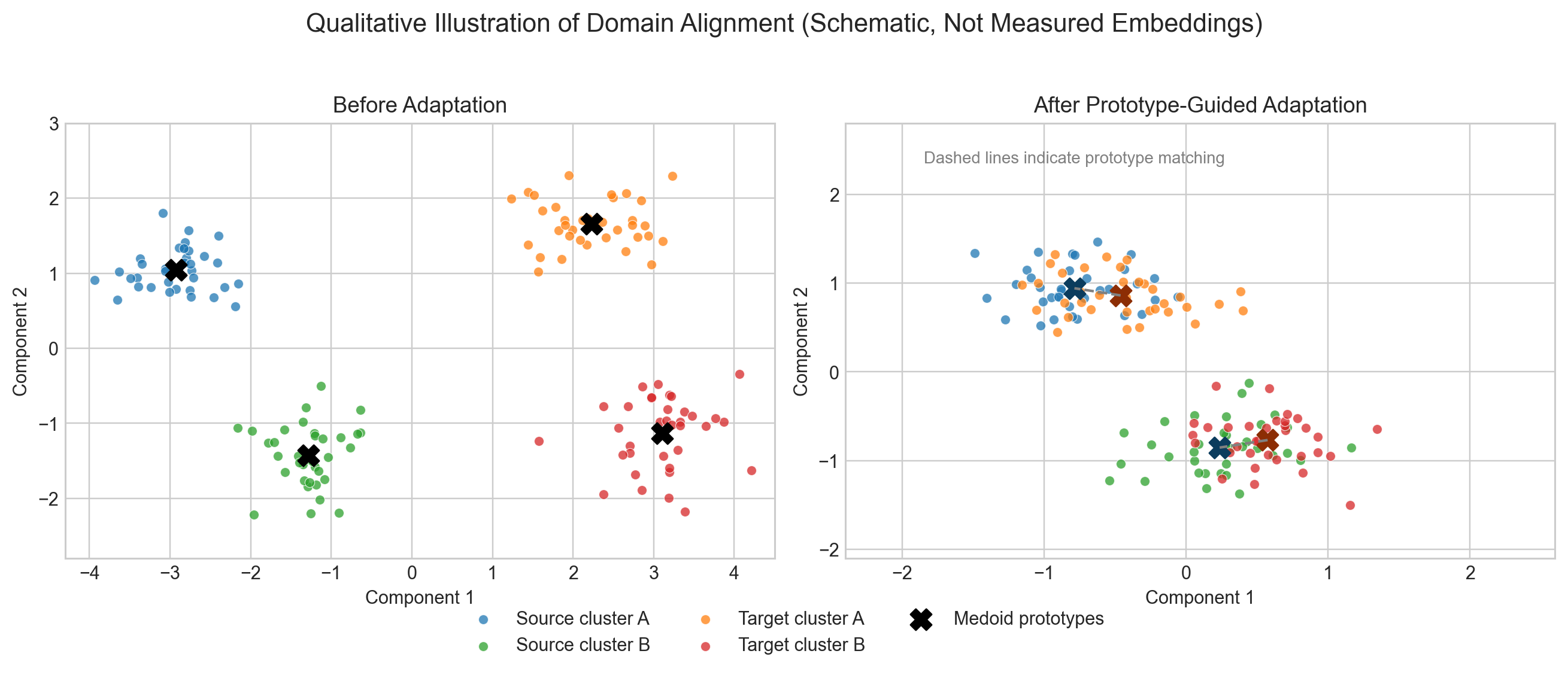}
    \caption{Qualitative illustration of domain alignment behavior. This figure is schematic and is included for interpretation only; it is not a measured t-SNE or UMAP embedding from the current experiment package.}
    \label{fig:qualitative_alignment}
\end{figure}

Because the current manuscript package does not include stored embedding outputs for direct t-SNE or UMAP visualization, we provide a schematic qualitative illustration in Fig.~\ref{fig:qualitative_alignment}. The figure does not represent measured coordinates. Instead, it summarizes the intended effect of MPA: source and target samples are initially separated in the shared feature space, while prototype-guided adaptation brings corresponding operational regions closer together through medoid matching.

\subsection{Discussion}

Three observations emerge from the experiments. First, average performance is a more informative indicator than isolated task-wise peaks for cross-plant deployment, and MPA shows the strongest average behavior. Second, transfer direction matters: source plants with richer operational semantics generally support better adaptation. Third, summarizing each domain through medoid prototypes appears to stabilize adaptation when industrial traffic is noisy, imbalanced, or heterogeneous. Together, these observations support the idea that local structural anchors are a useful complement to conventional domain adaptation in industrial cybersecurity.

\section{Conclusion}

This paper presented a medoid prototype alignment framework for cross-plant unknown attack detection in Industrial Control Systems. By combining shared-space compression, robust medoid extraction, and prototype-calibrated transfer learning, the proposed method provides a structurally grounded alternative to direct global alignment. The experimental analysis shows that the method achieves the best average performance across four cross-domain industrial tasks and remains more stable under difficult transfer directions. These results suggest that prototype-guided adaptation is a promising direction for practical industrial intrusion detection under scarce labels and evolving deployments. Future work will extend this idea to open-set labeling, online adaptation, and real-time industrial monitoring.

{
\bibliographystyle{plain}
\bibliography{sample}

@inproceedings{wang2024uvmap,
  title={UVMap-ID: A Controllable and Personalized UV Map Generative Model},
  author={Wang, Weijie and Zhang, Jichao and Liu, Chang and Li, Xia and Xu, Xingqian and Shi, Humphrey and Sebe, Nicu and Lepri, Bruno},
  booktitle={ACM MM},
  pages={10725--10734},
  year={2024}
}

@misc{li2025freeinsert,
      title={FreeInsert: Disentangled Text-Guided Object Insertion in 3D Gaussian Scene without Spatial Priors}, 
      author={Chenxi Li and Weijie Wang and Qiang Li and Bruno Lepri and Nicu Sebe and Weizhi Nie},
      booktitle={ACM MM},
      year={2025},
}

@article{nie2024t2td,
  title={T2TD: Text-3D generation model based on prior knowledge guidance},
  author={Nie, Weizhi and Chen, Ruidong and Wang, Weijie and Lepri, Bruno and Sebe, Nicu},
  journal={IEEE TPAMI},
  year={2024},
  publisher={IEEE}
}

@INPROCEEDINGS{11125558,
  author={Wang, Weijie and Mei, Guofeng and Zhang, Jian and Sebe, Nicu and Lepri, Bruno and Poiesi, Fabio},
  booktitle={3DV}, 
  title={Fully-Geometric Cross-Attention for Point Cloud Registration}, 
  year={2025},
  }

@inproceedings{wang2025fully,
  title={Fully-geometric cross-attention for point cloud registration},
  author={Wang, Weijie and Mei, Guofeng and Zhang, Jian and Sebe, Nicu and Lepri, Bruno and Poiesi, Fabio},
  booktitle={3DV},
  year={2025},
  organization={IEEE}
}

@article{wang2023turn,
  title={Turn fake into real: Adversarial head turn attacks against deepfake detection},
  author={Wang, Weijie and Zhao, Zhengyu and Sebe, Nicu and Lepri, Bruno},
  journal={arXiv preprint arXiv:2309.01104},
  year={2023}
}

@inproceedings{mei2023udpreg,
  author    = {Guofeng Mei and Hao Tang and Xiaoshui Huang and Weijie Wang and Juan Liu and Jian Zhang and Luc Van Gool and Qiang Wu},
  title     = {Unsupervised Deep Probabilistic Approach for Partial Point Cloud Registration},
  booktitle = {Proceedings of the IEEE/CVF Conference on Computer Vision and Pattern Recognition (CVPR)},
  year      = {2023}
}

@inproceedings{wang2023diga,
  author    = {Wei Wang and Zhun Zhong and Weijie Wang and Xi Chen and Charles Ling and Boyu Wang and Nicu Sebe},
  title     = {Dynamically Instance-Guided Adaptation: A Backward-Free Approach for Test-Time Domain Adaptive Semantic Segmentation},
  booktitle = {Proceedings of the IEEE/CVF Conference on Computer Vision and Pattern Recognition (CVPR)},
  pages     = {24090--24099},
  year      = {2023}
}

@article{chen2024llmgeo,
  author  = {Yuxing Chen and Weijie Wang and Sylvain Lobry and Camille Kurtz},
  title   = {An LLM Agent for Automatic Geospatial Data Analysis},
  journal = {arXiv preprint arXiv:2410.18792},
  year    = {2024}
}

@article{nie2019hgan,
  author  = {Weizhi Nie and Weijie Wang and Anan Liu and Jie Nie and Yuxuan Su},
  title   = {HGAN: Holistic Generative Adversarial Networks for Two-Dimensional Image-Based Three-Dimensional Object Retrieval},
  journal = {ACM Transactions on Multimedia Computing, Communications, and Applications},
  volume  = {15},
  number  = {4},
  pages   = {1--24},
  year    = {2019},
  doi     = {10.1145/3344684}
}

@inproceedings{ren2024pointcmae,
  author    = {Bin Ren and Guofeng Mei and Danda Pani Paudel and Weijie Wang and Yawei Li and Mengyuan Liu and Rita Cucchiara and Luc Van Gool and Nicu Sebe},
  title     = {Bringing Masked Autoencoders Explicit Contrastive Properties for Point Cloud Self-Supervised Learning},
  booktitle = {Proceedings of the Asian Conference on Computer Vision (ACCV)},
  year      = {2024},
  doi       = {10.1007/978-981-96-0963-5_4}
}

@article{wang2024ssdag,
  author  = {Xuquan Wang and Feng Zhang and Kai Zhang and Weijie Wang and Xiong Dun and Jiande Sun},
  title   = {Learning Spatial-Spectral Dual Adaptive Graph Embedding for Multispectral and Hyperspectral Image Fusion},
  journal = {Pattern Recognition},
  volume  = {151},
  pages   = {110365},
  year    = {2024}
}

@inproceedings{nie2019characteristic,
  author    = {Weizhi Nie and Weijie Wang and Anan Liu and Chuang Chen},
  title     = {Characteristic Views Extraction Modal Based-on Deep Reinforcement Learning for 3D Model Retrieval},
  booktitle = {2019 IEEE International Conference on Image Processing (ICIP)},
  pages     = {2389--2393},
  year      = {2019},
  doi       = {10.1109/ICIP.2019.8803343}
}

@article{chang2024shapekg,
  author  = {Rihao Chang and Yongtao Ma and Tong Hao and Weijie Wang and Weizhi Nie},
  title   = {3D Shape Knowledge Graph for Cross-Domain 3D Shape Retrieval},
  journal = {CAAI Transactions on Intelligence Technology},
  volume  = {9},
  number  = {5},
  pages   = {1199--1216},
  year    = {2024},
  doi     = {10.1049/cit2.12326}
}

@article{wang2022fer,
  author  = {Weijie Wang and Nicu Sebe and Bruno Lepri},
  title   = {Rethinking the Learning Paradigm for Facial Expression Recognition},
  journal = {arXiv preprint arXiv:2209.15402},
  year    = {2022}
}

@article{liang2020lstm3d,
  author  = {Qi Liang and Ning Xu and Weijie Wang and Xingjian Long},
  title   = {Multimodal Information Fusion Based on LSTM for 3D Model Retrieval},
  journal = {Multimedia Tools and Applications},
  volume  = {79},
  number  = {45--46},
  pages   = {33943--33956},
  year    = {2020},
  doi     = {10.1007/s11042-020-08817-6}
}

@article{xu2025scmllm,
  author  = {Zibo Xu and Qiang Li and Weizhi Nie and Weijie Wang and Anan Liu},
  title   = {Structure Causal Models and LLMs Integration in Medical Visual Question Answering},
  journal = {IEEE Transactions on Medical Imaging},
  volume  = {44},
  number  = {8},
  pages   = {3476--3489},
  year    = {2025},
  doi     = {10.1109/TMI.2025.3564320}
}

@article{wang2023zeroreg,
  author  = {Weijie Wang and Wenqi Ren and Guofeng Mei and Bin Ren and Xiaoshui Huang and Fabio Poiesi and Nicu Sebe and Bruno Lepri},
  title   = {ZeroReg: Zero-Shot Point Cloud Registration with Foundation Models},
  journal = {arXiv preprint arXiv:2312.03032},
  year    = {2023}
}

@article{ma2023rfrwwanet,
  author  = {Mingrui Ma and Tao Wang and Liyuan Song and Weijie Wang and Guixia Liu},
  title   = {RFR-WWANet: Weighted Window Attention-Based Recovery Feature Resolution Network for Unsupervised Image Registration},
  journal = {Pattern Recognition},
  year    = {2023}
}

@article{wang2025mbtpolyp,
  author  = {Tao Wang and Weijie Wang and Fausto Giunchiglia and Fengzhi Zhao and Ye Zhang and Duo Yu and Guixia Liu},
  title   = {MBT-Polyp: A New Multi-Branch Memory-Augmented Transformer for Polyp Segmentation},
  journal = {Image and Vision Computing},
  volume  = {163},
  pages   = {105747},
  year    = {2025},
  doi     = {10.1016/j.imavis.2025.105747}
}

@article{ma2024llmmorph,
  author  = {Mingrui Ma and Weijie Wang and Jie Ning and Jianfeng He and Nicu Sebe and Bruno Lepri},
  title   = {Large Language Models for Multimodal Deformable Image Registration},
  journal = {arXiv preprint arXiv:2408.10703},
  year    = {2024}
}

@inproceedings{wang2024uhrmlp,
  author    = {Tao Wang and Kai Zhang and Weijie Wang and Mingrui Ma and Ye Zhang and He Zhao and Guixia Liu},
  title     = {U-HRMLP: Refining Segmentation Boundaries in Histopathology Images},
  booktitle = {2024 IEEE International Symposium on Biomedical Imaging (ISBI)},
  pages     = {1--5},
  year      = {2024},
  doi       = {10.1109/ISBI56570.2024.10635255}
}

@article{gao2025pointpc,
  author  = {Xuesong Gao and Chuanqi Jiao and Ruidong Chen and Weijie Wang and Weizhi Nie},
  title   = {Point-PC: Point Cloud Completion Guided by Prior Knowledge via Causal Inference},
  journal = {CAAI Transactions on Intelligence Technology},
  year    = {2025},
  doi     = {10.1049/cit2.12379}
}

@article{chang2025organagents,
  author  = {Rihao Chang and Hongbo Jiao and Weizhi Nie and Huijie Guo and Kai Xie and Zihan Wu and Lin Zhao and Yutong Bai and Yongtao Ma and Lijuan Wang and others},
  title   = {Organ-Agents: Virtual Human Physiology Simulator via LLMs},
  journal = {arXiv preprint arXiv:2508.14357},
  year    = {2025}
}

@article{rigo2026pocidiff,
  author  = {A. Rigo and L. Stornaiuolo and Weijie Wang and M. Martino and Bruno Lepri and Nicu Sebe and Weizhi Nie},
  title   = {POCI-Diff: Position Objects Consistently and Interactively with 3D-Layout Guided Diffusion},
  journal = {arXiv preprint arXiv:2601.14056},
  year    = {2026}
}

@article{umer2022mlics,
  author  = {Umer, Muhammad Azmi and Junejo, Khurum Nazir and Jilani, Muhammad Taha and Mathur, Aditya P.},
  title   = {Machine Learning for Intrusion Detection in Industrial Control Systems: Applications, Challenges, and Recommendations},
  journal = {International Journal of Critical Infrastructure Protection},
  volume  = {38},
  pages   = {100516},
  year    = {2022},
  doi     = {10.1016/j.ijcip.2022.100516}
}

@article{kheddar2023dtl,
  author  = {Kheddar, Hamza and Himeur, Yassine and Awad, Ali Ismail},
  title   = {Deep Transfer Learning for Intrusion Detection in Industrial Control Networks: A Comprehensive Review},
  journal = {Journal of Network and Computer Applications},
  volume  = {220},
  pages   = {103760},
  year    = {2023},
  doi     = {10.1016/j.jnca.2023.103760}
}

@article{gauthama2021icsml,
  author  = {Gauthama Raman, M. R. and Ahmed, Chuadhry Mujeeb and Mathur, Aditya},
  title   = {Machine Learning for Intrusion Detection in Industrial Control Systems: Challenges and Lessons from Experimental Evaluation},
  journal = {Cybersecurity},
  volume  = {4},
  number  = {1},
  pages   = {27},
  year    = {2021},
  doi     = {10.1186/s42400-021-00095-5}
}

@article{ganin2016dann,
  author  = {Ganin, Yaroslav and Ustinova, Evgeniya and Ajakan, Hana and Germain, Pascal and Larochelle, Hugo and Laviolette, Fran{\c{c}}ois and Marchand, Mario and Lempitsky, Victor},
  title   = {Domain-Adversarial Training of Neural Networks},
  journal = {Journal of Machine Learning Research},
  volume  = {17},
  number  = {59},
  pages   = {1--35},
  year    = {2016},
  url     = {https://jmlr.org/papers/v17/15-239.html}
}

@article{zhao2019unknownattack,
  author  = {Zhao, Juan and Shetty, Sachin and Pan, Jan Wei and Kamhoua, Charles and Kwiat, Kevin},
  title   = {Transfer Learning for Detecting Unknown Network Attacks},
  journal = {EURASIP Journal on Information Security},
  volume  = {2019},
  number  = {1},
  pages   = {1},
  year    = {2019},
  doi     = {10.1186/s13635-019-0084-4}
}

@article{kravchik2018icscnn,
  author  = {Kravchik, Moshe and Shabtai, Asaf},
  title   = {Detecting Cyberattacks in Industrial Control Systems Using Convolutional Neural Networks},
  journal = {arXiv preprint arXiv:1806.08110},
  year    = {2018},
  url     = {https://arxiv.org/abs/1806.08110}
}

@incollection{kaufman1990pam,
  author    = {Kaufman, Leonard and Rousseeuw, Peter J.},
  title     = {Partitioning Around Medoids (Program PAM)},
  booktitle = {Finding Groups in Data: An Introduction to Cluster Analysis},
  publisher = {John Wiley \& Sons},
  address   = {New York},
  pages     = {68--125},
  year      = {1990},
  doi       = {10.1002/9780470316801.ch2}
}

@article{pearson1901pca,
  author  = {Pearson, Karl},
  title   = {On Lines and Planes of Closest Fit to Systems of Points in Space},
  journal = {The London, Edinburgh, and Dublin Philosophical Magazine and Journal of Science},
  volume  = {2},
  number  = {11},
  pages   = {559--572},
  year    = {1901},
  doi     = {10.1080/14786440109462720}
}

@book{jolliffe2002pca,
  author    = {Jolliffe, Ian T.},
  title     = {Principal Component Analysis},
  edition   = {2},
  publisher = {Springer},
  address   = {New York},
  year      = {2002},
  isbn      = {9780387954424}
}

@article{mushtaq2018pam,
  author  = {Mushtaq, Haroon and Khan, S. U. and Jan, M. A. and Ullah, A. and Khattak, H. A.},
  title   = {A Parallel Architecture for the Partitioning Around Medoids (PAM) Algorithm},
  journal = {Sensors},
  volume  = {18},
  number  = {12},
  pages   = {4129},
  year    = {2018},
  doi     = {10.3390/s18124129}
}

@article{bendavid2010domain,
  author  = {Ben-David, Shai and Blitzer, John and Crammer, Koby and Pereira, Fernando},
  title   = {A Theory of Learning from Different Domains},
  journal = {Machine Learning},
  volume  = {79},
  number  = {1--2},
  pages   = {151--175},
  year    = {2010},
  doi     = {10.1007/s10994-009-5152-4}
}

@inproceedings{pan2009tca,
  author    = {Pan, Sinno Jialin and Tsang, Ivor W. and Kwok, James T. and Yang, Qiang},
  title     = {Domain Adaptation via Transfer Component Analysis},
  booktitle = {Proceedings of the 21st International Joint Conference on Artificial Intelligence (IJCAI)},
  pages     = {1187--1192},
  year      = {2009}
}

@article{chen2023crossdomain,
  author  = {Chen, Y. and Su, S. and Yu, D. and He, H. and Wang, X. and Ma, Y. and Guo, H.},
  title   = {Cross-Domain Industrial Intrusion Detection Deep Model Trained With Imbalanced Data},
  journal = {IEEE Internet of Things Journal},
  volume  = {10},
  pages   = {584--596},
  year    = {2023}
}

@ARTICLE{10684147,
  author={Nie, Weizhi and Chen, Ruidong and Wang, Weijie and Lepri, Bruno and Sebe, Nicu},
  journal={IEEE Transactions on Pattern Analysis and Machine Intelligence}, 
  title={T2TD: Text-3D Generation Model Based on Prior Knowledge Guidance}, 
  year={2025},
  volume={47},
  number={1},
  pages={172-189},
  doi={10.1109/TPAMI.2024.3463753}}

@article{wang2026poinit,
  title={PoInit-of-View: Poisoning Initialization of Views Transfers Across Multiple 3D Reconstruction Systems},
  author={Wang, Weijie and Xing, Songlong and Zhao, Zhengyu and Sebe, Nicu and Lepri, Bruno},
  journal={arXiv preprint arXiv:2604.16540},
  year={2026}
}

@article{xing2026finetune,
  title={Finetune Like You Pretrain: Boosting Zero-shot Adversarial Robustness in Vision-language Models},
  author={Xing, Songlong and Wang, Weijie and Zhao, Zhengyu and Gu, Jindong and Torr, Philip and Sebe, Nicu},
  journal={arXiv preprint arXiv:2604.11576},
  year={2026}
}

@article{gao2023point,
  title={Point cloud completion guided by prior knowledge via causal inference},
  author={Gao, Songxue and Jiao, Chuanqi and Chen, Ruidong and Wang, Weijie and Nie, Weizhi},
  journal={arXiv preprint arXiv:2305.17770},
  year={2023}
}
}

\end{document}